\RequirePackage[l2tabu, orthodox]{nag}
\RequirePackage{snapshot}

\documentclass[11pt,onecolumn]{article}

\makeatletter
\if@twocolumn
  \usepackage[dvips,letterpaper,top=0.5in, bottom=0.5in, left=0.75in, right=0.5in,includefoot,heightrounded]{geometry}
\else
  \usepackage[dvips,letterpaper,margin=1in,includefoot,heightrounded]{geometry}
\fi

\usepackage{srcltx}

\usepackage{amsmath}
\usepackage{amssymb,amsfonts}

\usepackage{abstract}

\usepackage{epsfig}
\usepackage[usenames,dvipsnames]{color}
\usepackage[usenames,dvipsnames]{xcolor}
\usepackage{subfigure}

\usepackage{booktabs}

\usepackage{setspace}
\usepackage{flushend}
\usepackage{multicol}

\usepackage{cite}
\usepackage{url}
\usepackage[normalem]{ulem}

\usepackage{enumerate}

\usepackage{hyperref}

\usepackage{lipsum}

\usepackage[sc,tiny,center]{titlesec}

\makeatletter

\newenvironment{rsmallmatrix}{\null\,\vcenter\bgroup
  \Let@\restore@math@cr\default@tag
  \baselineskip6\ex@ \lineskip1.5\ex@ \lineskiplimit\lineskip
  \ialign\bgroup\hfil$\m@th\scriptstyle##$&&\thickspace\hfil
  $\m@th\scriptstyle##$\crcr
}{%
  \crcr\egroup\egroup\,%
}
\makeatother

\title{%
Multi-beam 4~GHz Microwave Apertures Using Current-Mode DFT Approximation on 65~nm CMOS
}

\author{%
Viduneth~Ariyarathna$^\ast$
\and
Sunera~Kulasekera$^\ast$
\and
Arjuna~Madanayake$^\ast$
\and
Kye-Shin Lee%
\thanks{%
V. Ariyarathna,
S. Kulasekera,
A. Madanayake,
and K.-S. Lee are with the
Department of Electrical and Computer Engineering, University of Akron, Akron, OH, USA 44325-3904.
E-mail:
\protect\url{arjuna@uakron.edu}}
\and
Dora~Suarez%
\thanks{%
D. Suarez is with the
Signal Processing Group, Departamento de Estat\'{\i}stica, UFPE, Brazil.}
\and
R.~J.~Cintra%
\thanks{%
R.~J.~Cintra is with
the Signal Processing Group,
Departamento de Estat\'{\i}stica,
Universidade Federal de Pernambuco;
Equipe Cairn, IRISA/INRIA, Universit\'e de Rennes 1, Rennes, France;
LIRIS,
and
Institut National des Sciences Appliqu\'ees,
Lyon, France.
E-mail:
\protect\url{rjdsc@stat.ufpe.org}
}
\and
F\'abio~M.~Bayer%
\thanks{%
F.~M.~Bayer is with the
Departamento de Estat\'{\i}stica,
Universidade Federal de Santa Maria, Brazil.
}
\and
Leonid Belostotski%
\thanks{%
L. Belostotski is with the
Department of Electrical and Computer Engineering, University of Calgary, Calgary, AB T2N 1N4, Canada.
}
}

\date{}

\newcommand{\myabstract}{%
A current-mode CMOS design is proposed for realizing receive mode multi-beams in the analog domain using a novel DFT approximation. High-bandwidth CMOS RF transistors are employed in low-voltage current mirrors  to achieve bandwidths exceeding 4~GHz with good beam fidelity. Current mirrors realize
the coefficients of the
considered DFT approximation, which take simple values in $\{0, \pm1, \pm2\}$ only. This allows high bandwidths realizations using simple circuitry without needing phase-shifters or delays. The proposed design is used as a method to efficiently achieve spatial discrete Fourier transform operation across a ULA to obtain multiple simultaneous RF beams. An example using 1.2~V current-mode approximate DFT on 65~nm CMOS, with BSIM4 models from the RF kit, show potential operation up to 4~GHz with eight independent aperture beams.
}

\newcommand{\mykeywords}{%
Analog, arrays, beamforming, aperture, multibeam.
}

\begin{document}

\makeatletter
\if@twocolumn

\twocolumn[%
  \maketitle
  \begin{onecolabstract}
    \myabstract
  \end{onecolabstract}
  \begin{center}
    \small
    \textbf{Keywords}
    \\\medskip
    \mykeywords
  \end{center}
  \bigskip
]
\saythanks

\else

  \maketitle
  \begin{abstract}
    \myabstract
  \end{abstract}
  \begin{center}
    \small
    \textbf{Keywords}
    \\\medskip
    \mykeywords
  \end{center}
  \bigskip
  \onehalfspacing
\fi

\section{Introduction}
The formation of multiple orthogonal radio-frequency (RF) beams are a quintessential example of antenna arrays~\cite{patnika,kalia}. There are applications for multi-beam arrays, such as wireless communications, radar, radio astronomy, and space imaging. Alternative to
phased-arrays~\cite{eric} that achieve a  single steerable beam, multiple simultaneous
RF beams---directed at fixed directions---are achieved using a \emph{spatial discrete Fourier transform (DFT) operation} across a uniformly-spaced array~\cite{Ellingson, ONR}.
In digital radar, multi-beams are achieved by digitizing the intermediate frequency (IF) signals from each element receiver, which are connected to the array. Sampling is followed by an application
of the $N$-point spatial DFT
computed via a
fast Fourier transform (FFT) on a per-frame basis~\cite{Ellingson, ONR}.

\begin{figure}
\centering
\scalebox{0.6}{\input{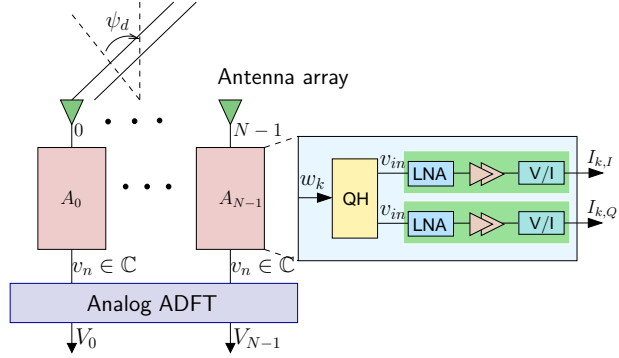}}
\caption{Analog RF 8-beam aperture using a spatial 8-point DFT approximation.}
\label{ADFT_1}
\end{figure}

The use of an 8-point approximate DFT, implemented by means of a fast algorithm~\cite{Suarez2014},
where DFT-beams have been closely emulated using a matrix of Gaussian integer weights, allows multi-beams using relatively simple active RF circuits.
Approximate transformations are
linear transformations of low computational cost
offering close results to that from the exact transformation.
The realization of multi-beams in analog
by mean of a DFT approximation
using microwave circuits exploits the high-bandwidth  of current-mode CMOS integrated circuits.
The approximate DFT achieves RF beams that are nearly identical to DFT-beams albeit without a Butler Matrix. That is, the analog  circuit becomes quite simple to design owing to the use of current mirrors having weights
over the set $\mathcal{P} = \{0, \pm1, \pm2\}$ only.
We show that such an approximation can be efficiently realized at 4~GHz or more of bandwidth using analog IC designs. Fig.~\ref{ADFT_1} shows the overview of the aperture array, where $N$ elements makeup a uniformly-spaced linear array (ULA) with spacing $\Delta x$. The elements (e.g., Vivaldi or spiral antennas) are amplified and quadrature down-converted or fed through a quadrature hybrid (QH)  to achieve complex inputs for the
8-point DFT approximation.
\section{8-point Approximate DFT Multi-Beam Matrix}
\label{DFT}
Recall that a DFT is a discrete orthogonal transformation
that transforms an input vector $\mathbf{v}= \begin{bmatrix}v_{0}(t) & v_1(t) & \cdots &v_{N}(t)\end{bmatrix}^\top$
to an output vector with $N$ spectral coefficients,
denoted by
$\mathbf{V}=\begin{bmatrix}V_{0}(t) & V_1(t) & \cdots & V_{N}(t) \end{bmatrix}^\top$~\cite{balhut2010}, each corresponding to a far-field RF beam for $N$ element arrays when $v_k(t)=x_I(t)+jx_Q(t)\in\mathbb{C}$ consist of in-phase $(x_I(t))$ and quadrature $(x_Q(t))$ antenna feeds from a QH component per array element, and where $
V_{k}(t)
=
\sum_{n=0}^{N-1}
v_{n}(t)
\cdot
\omega_N^{kn}
,
\quad
k=0,1,\ldots,N-1,
$
where
$j=\sqrt{-1}$
and
$\omega_N = \exp\left\{-\frac{2\pi j}{N}\right\}$
is the $N$th root of unity.
In matrix form,
$\mathbf{V}
=
\mathbf{F}_{N}
\cdot
\mathbf{v}$
and
$\mathbf{v}
=
\mathbf{F}_{N}^{-1}
\cdot
\mathbf{V}
=
\frac{1}{N}
\cdot
\mathbf{F}_{N}^{*}
\cdot
\mathbf{V}
$,
where $\mathbf{F}_{N}$
is the DFT transformation matrix whose
$(i,k)$th element is
$f_{i,k}=\omega_N^{(i-1)(k-1)}$,
for $i,k=1,2,\ldots,N$
and the superscript~${}^\ast$
denotes the transposed conjugation (Hermitian).
Classical radar apertures obtain multi-beams using a Butler Matrix realization of the DFT.
For the derivation of
DFT approximations,
we consider the set $\mathcal{Q}
=
\Big\{
z \in \mathbb{C}
\,
:
\,
\Re\{z \} \in \mathcal{P}
\land
\Im \{z \} \in \mathcal{P}
\Big \}
$
for the
matrix entries of
the
DFT approximation matrices.
Parametric optimization,
which minimizes
the Frobenius norm,
over the set $\mathcal{Q}$
leads to
an 8-point DFT approximation
having near-orthogonality
and low circuit complexity.
The optimal
elements
for the parametric approximation of $\mathbf{F}_8$
are
$1$,
$(1-j)/2$,
and
$-j$.
Thus,
the resulting
approximate DFT matrix contains
only Gaussian integer entries:
\begin{align*}
\hat{\mathbf{F}}_8
=
\frac{1}{2}
\cdot
\left[
\begin{rsmallmatrix}
  2 & 2 & 2 & 2 & 2 & 2 & 2 & 2 \\
  2 & 1-j & -2j & -1-j & -2 & -1+j & 2j & 1+j \\
  2 & -2j & -2 & 2j & 2 & -2j & -2 & 2j \\
  2 & -1-j & 2j & 1-j & -2 & 1+j & -2j & -1+j \\
  2 & -2 & 2 & -2 & 2 & -2 & 2 & -2 \\
  2 & -1+j & -2j & 1+j & -2 & 1-j & 2j & -1-j \\
  2 & 2j & -2 & -2j & 2 & 2j & -2 & -2j \\
  2 & 1+j & 2j & -1+j & -2 & -1-j & -2j & 1-j \\
\end{rsmallmatrix}
\right]
.
\end{align*}

The above 8-point approximate DFT
matrix~$\hat{\mathbf{F}}_8$
preserves the symmetry of the DFT
and has null multiplicative complexity---a salient property that allows
realization using current-mode circuits having integer multiplications of current values that map neatly into multiples of identical 1:1 mirrors on an analog IC.
Let $\mathbf{I}_n$
be the
identity matrix of order $n$
and
$\mathbf{B}_n =
\left[\begin{rsmallmatrix}1 & 1\\ 1 & -1 \end{rsmallmatrix}\right]
\otimes
\mathbf{I}_{n/2}$,
where $\otimes$ denotes the Kronecker product.
Matrix factorization techniques
leads to an algorithm for mapping to analog circuits~\cite{balhut2010}.
In particular,
$\mathbf{\hat{F}}_{8}$ admits the following factorization:
\begin{align*}
\mathbf{\hat{F}}_{8}
=&
\mathbf{P}
\times
\operatorname{diag}
\big(
\mathbf{I}_2,\mathbf{A}_1,\mathbf{A}_3
\big)
\times
\mathbf{D}_2
\times
\operatorname{diag}
\big(
\mathbf{B}_2,\mathbf{I}_2,\mathbf{A}_4
\big)
\\
&
\times
\mathbf{D}_1
\times
\operatorname{diag}
\big(
\mathbf{B}_{4},\mathbf{A}_2
\big)
\times
\mathbf{B}_8
,
\end{align*}
where
$
\mathbf{A}_{1}=
\left[
\begin{rsmallmatrix}
1	&	-1\\
1	&	1
\end{rsmallmatrix}
\right]
$,
$
\mathbf{A}_{2}=
\left[
\begin{rsmallmatrix}
1 & & & \\
  & 1	&	& 1\\
  & &	1 & \\
& 1 & & -1
\end{rsmallmatrix}
\right]
$,
$
\mathbf{A}_{3}=
\left[
\begin{rsmallmatrix}
1	&-1	& & \\
&	 & -1 & 1 \\
1 & 1 &  & \\
 & & 1& 1
\end{rsmallmatrix}
\right]
$,
$
\mathbf{A}_{4}=
\left[
\begin{rsmallmatrix}
1	&	& &1 \\
&	1 & 1 & \\
&1 & -1 &\\
1 & & & -1
\end{rsmallmatrix}
\right]
$,
$\mathbf{D}_1=\operatorname{diag}
(
\begin{rsmallmatrix}
1, & 1, & 1, & 1, & 1, & 1/2, & 1, & 1/2
\end{rsmallmatrix}
)$,
$\mathbf{D}_2=\operatorname{diag}
(
\begin{rsmallmatrix}
1, & 1, & 1, & j, & 1, & j, & j, & 1
\end{rsmallmatrix}
)$,
$
\mathbf{P}
=
\left[
\begin{rsmallmatrix}
\mathbf{e}_1&\big|\,
\mathbf{e}_5&\big|\,
\mathbf{e}_3&\big|\,
\mathbf{e}_6&\big|\,
\mathbf{e}_2&\big|\,
\mathbf{e}_8&\big|\,
\mathbf{e}_4&\big|\,
\mathbf{e}_7
\end{rsmallmatrix}
\right]
^\top
$
is a
permutation matrix,
and
$\mathbf{e}_i$ is
the 8-point column vector with
element~1 at the $i$th position
and 0~elsewhere.

\section{8-beam 4~GHz DFT Approximation in 65~nm CMOS}

\subsection{Antenna Front-End}

\begin{figure}
\centering

\subfigure[]{\includegraphics[width=0.5\columnwidth]{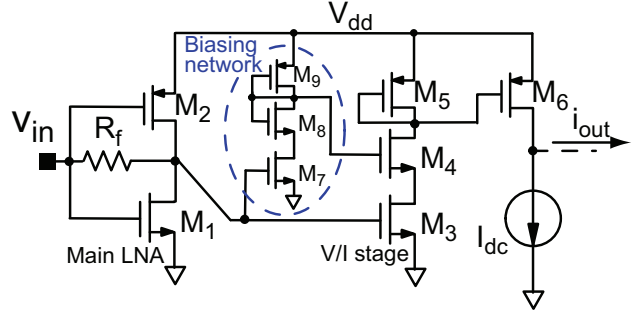}}

\subfigure[]{\includegraphics[width=0.5\columnwidth]{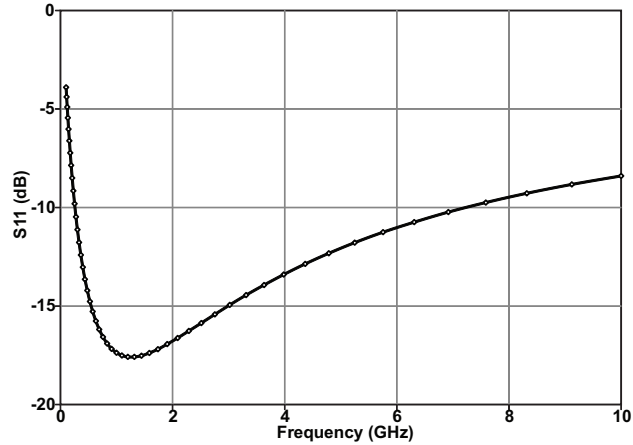}}

\caption{\label{fig:LNA} 65~nm 1.2~V CMOS LNA with current output $i_{out}$ (top);
amplifier $|S_{11}|$[dB] better than $-10$~dB up to 7~GHz (bottom).}

\end{figure}

A conventional inverter-based shunt feedback LNA, shown in Fig.~\ref{fig:LNA}(a),
is employed.
The input signal,
$v_\text{in}$, from each antenna is applied
to the gates of $M_{1}$ and $M_{2}$, forming the main LNA, whose
$S_{11}$, displayed in Fig.~\ref{fig:LNA}(b), is set with $R_{f}$
and the transconductances of $M_{1,2}$. The cascoded transistors
$M_{3,4}$ are used to increase the impedance at the drain of $M_{5}$
so that most of the small-signal current flows into the $M_{5,6}$
current mirror. The circuit consisting of $M_\text{3-6}$ forms a voltage-to-current
(V/I) conversion stage. A DC current source $I_\text{dc}$ reduces
the~DC component of the $i_\text{out}$ current.
The gate bias for $M_{4}$
is provided by a biasing network consisting of $M_\text{7-9}$, which are
scaled versions of $M_\text{3-5}$. The LNA draws $26.25\,\mbox{mA}$
from a 1.2~V~DC supply.

\begin{figure*}
\centering
\scalebox{.8}{\input{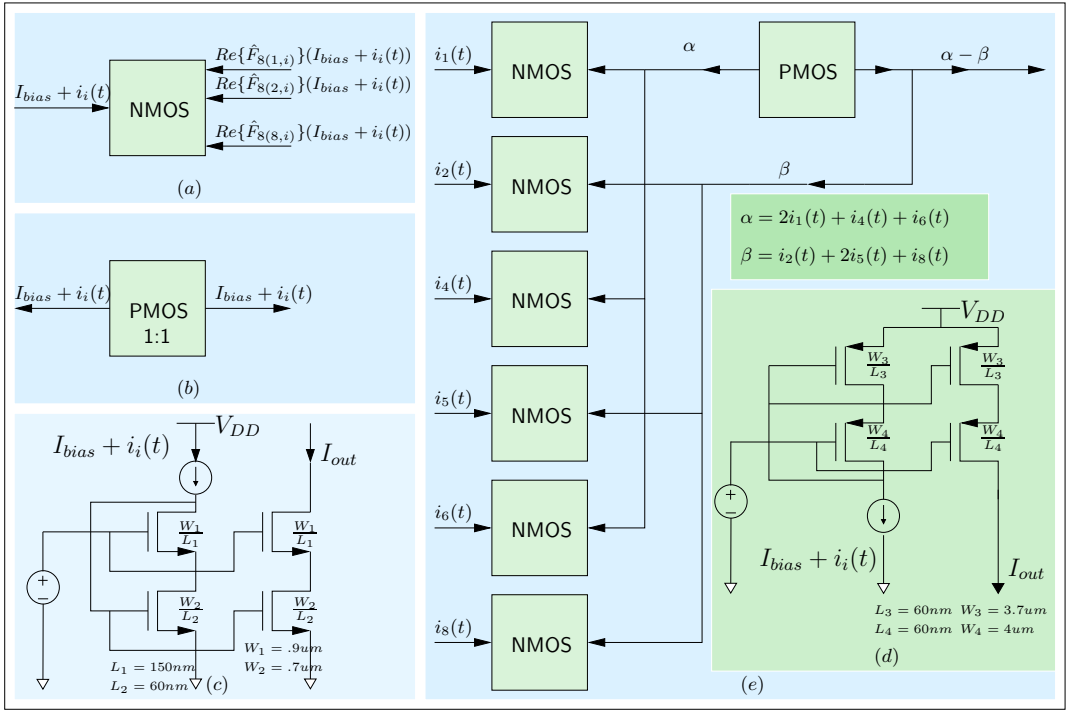}}
\caption{Current mode implementation of the 8-point DFT approximation using 65~nm CMOS RF NMOS/PMOS mirrors.}
\label{ADFT_2}
\end{figure*}

\begin{figure*}
\centering
\scalebox{0.9}{\input{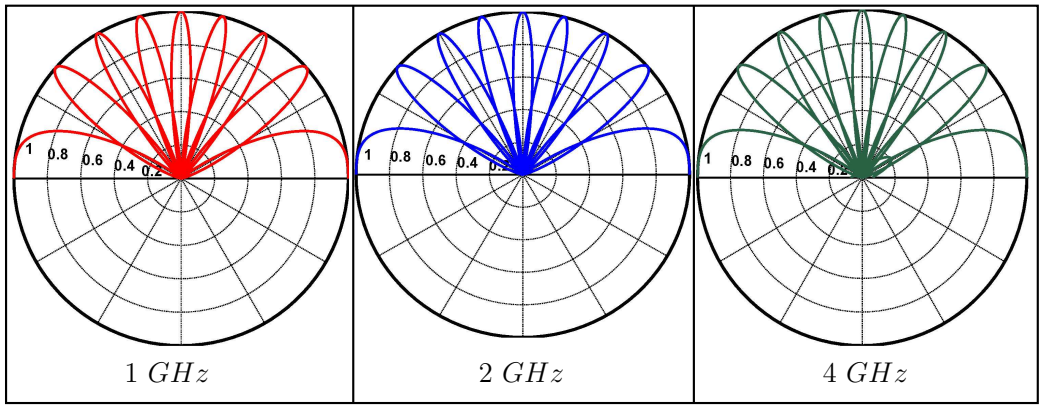}}
\caption{Polar response patterns for current outputs at $1$, $2$,
and~$4$~GHz.}
\label{ADFT_3}
\end{figure*}

\subsection{Current Mode Analog 8-point Approximate DFT}

The matrix $\hat{\mathbf{F}}_8$ is realized in current mode to increase operational bandwidth.
The signals from each LNA+secondary amplifier at the antennas (Fig.~\ref{ADFT_1}) are converted to output currents $i_\text{out}$. Eight such small-signal current outputs form the input signals for the current-mode realization of the discussed 8-point approximate
DFT. The real and imaginary components of the considered
approximate DFT matrix are realized separately. Fig.~\ref{ADFT_2}(a) shows a NMOS current copier used to realize one column of the DFT approximation matrix (real part). Fig.~\ref{ADFT_2}(b) shows the PMOS based current subtractor needed for negative valued entries of the DFT approximation matrix.
Example building blocks for the NMOS and PMOS current combiners in Fig.~\ref{ADFT_2}(a-b) are provided in Fig.~\ref{ADFT_2}(c-d), respectively. The DC bias current for the NMOS mirror in Fig.~\ref{ADFT_2}(c) is
$I_\text{bias}=100~\mu A$. An example of one row of the 8-point
approximate DFT (row~4) is shown in Fig.~\ref{ADFT_2}(e). All signal currents are assumed to be small (1--10\%) compared to DC bias currents. Each current mirror is designed using the low-voltage cascode topology~\cite{razavi}. The technology used is 65~nm GP CMOS, and all transistors are from the RF kit, with supply voltage 1.2~V. Simulations are in Cadence Spectre and employ BSIM4 RF transistor models.

\subsection{BSIM4 Array Patterns}

The Cadence designs were simulated at frequencies $f\in\{1,2, 4\}$ GHz.
The input currents were maintained at $2\mu A$ peak-to-peak.
The time-domain response was simulated, in steady state, and
the peak-to-peak values were noted.
The simulated small-signal output currents from the 8-point
DFT approximation outputs were used to compute the polar response patterns for each current mode circuit
for three frequencies in Fig.~\ref{ADFT_3}.
The polar patterns obtained from the BSIM4 models of the discussed
DFT approximation is very close to the expected ideal polar patterns
linked to the theoretical
8-point DFT approximation.
At higher frequencies, from the low-pass effects of the current mirrors due to dominant parasitic poles of the CMOS circuit, the pattern deviates noticeably (not shown here).
The 8-point approximate
DFT
provides far-field receive beams at directions
$90.0^\circ$,
$48.5^\circ$,
$30.0^\circ$,
$14.5^\circ$,
$0.0^\circ$,
$-14.5^\circ$,
$-30.0^\circ$,
and
$-48.5^\circ$ measured from the array.

\section{Conclusion}

The discussed
DFT approximation is a
numerical efficient method
for the approximate DFT evaluation,
and requires only
small Gaussian integer valued weights.
A multi-beam aperture algorithm and analog RF CMOS implementation for
the 8-point approximate DFT was proposed, designed, simulated and evaluated, for beamforming at bandwidths up to 4~GHz using ULAs of wideband elements. The analog approximate DFT circuit was designed using 65~nm GP CMOS employing RF transistors, with low-voltage current mirrors for maximum peak-peak swings and bandwidth. Cadence BSIM4 models  verify the aperture provides eight RF beams up to 4~GHz when supplied with 1.2~V~DC.

\bibliographystyle{IEEEtran}
\bibliography{dctref-clean}

\end{document}